\documentclass[conference]{IEEEtran}
\usepackage{cite}
\usepackage{amsmath,amssymb,amsfonts}
\usepackage{algorithmic}
\usepackage{graphicx}
\usepackage{textcomp}
\usepackage{xcolor}
\usepackage{epstopdf} %
\usepackage{url} %
\usepackage{mathtools}
\usepackage{gensymb}
\usepackage{booktabs}
\usepackage{svg}
\usepackage{subcaption}
\usepackage[nonumberlist]{glossaries}
\newacronym{wsn}{WSN}{Wireless Sensor Network}
\newacronym{mwsn}{MWSN}{Mobile Wireless Sensor Network}
\newacronym{iot}{IoT}{Internet of Things}
\newacronym{cso}{CSO}{Combined Sewer Overflow}
\newacronym{wgs}{WGS4.0}{WaterGridSense4.0}
\newacronym{senml}{SenML}{Sensor Markup Language}
\newacronym{lpwan}{LPWAN}{low-power wide-area network}
\newacronym{mqtt}{MQTT}{Message Queuing Telemetry Transport}
\newacronym{arima}{ARIMA}{AutoRegressive Integrated Moving Average}
\newacronym{lisa}{LISA}{Local Indicators of Spatial Association}
\newacronym{foss}{FOSS}{Free and Open Source Software}

\def\BibTeX{{\rm B\kern-.05em{\sc i\kern-.025em b}\kern-.08em
    T\kern-.1667em\lower.7ex\hbox{E}\kern-.125emX}}
\begin{document}

\title{A Scalable and Dependable Data Analytics Platform for Water Infrastructure Monitoring}

\author{
\IEEEauthorblockN{
    Felix Lorenz\IEEEauthorrefmark{1}, Morgan Geldenhuys\IEEEauthorrefmark{1}, Harald Sommer\IEEEauthorrefmark{2}, Frauke Jakobs\IEEEauthorrefmark{2},\\Carsten L\"uring\IEEEauthorrefmark{4}, Volker Skwarek\IEEEauthorrefmark{3}, Ilja Behnke\IEEEauthorrefmark{1} and Lauritz Thamsen\IEEEauthorrefmark{1}
\IEEEauthorblockA{
  \IEEEauthorrefmark{1}Technische Universit\"at Berlin, Germany, \{firstname.lastname\}@tu-berlin.de\\
}
\IEEEauthorblockA{
    \IEEEauthorrefmark{2}Ingenieurgesellschaft Prof. Dr. Sieker mbH, Germany, \{initial.lastname\}@sieker.de
}
\IEEEauthorblockA{
  \IEEEauthorrefmark{3}Hochschule für Angewandte Wissenschaften Hamburg, Germany, volker.skwarek@haw-hamburg.de
}
\IEEEauthorblockA{
  \IEEEauthorrefmark{4}WALTER tecyard GmbH \& Co. KG, Germany, carsten.luering@walter-tecyard.com
}
}
}

\IEEEoverridecommandlockouts

\maketitle

\IEEEpubidadjcol

\begin{abstract}
With weather becoming more extreme both in terms of longer dry periods and more severe rain events, municipal water networks are increasingly under pressure.
The effects include damages to the pipes, flash floods on the streets and combined sewer overflows.
Retrofitting underground infrastructure is very expensive, thus water infrastructure operators are increasingly looking to deploy IoT solutions that promise to alleviate the problems at a fraction of the cost.

In this paper, we report on preliminary results from an ongoing joint research project, specifically on the design and evaluation of its data analytics platform.
The overall system consists of energy-efficient sensor nodes that send their observations to a stream processing engine, which analyzes and enriches the data and transmits the results to a GIS-based frontend.
As the proposed solution is designed to monitor large and critical infrastructures of cities, several non-functional requirements such as scalability, responsiveness and dependability are factored into the system architecture. %
We present a scalable stream processing platform and its integration with the other components, as well as the algorithms used for data processing.
We discuss significant challenges and design decisions, introduce an efficient data enrichment procedure and present empirical results to validate the compliance with the target requirements.
The entire code for deploying our platform and running the data enrichment jobs is made publicly available with this paper.
\end{abstract}


\section{Introduction}\label{sec:intro}

Municipalities around the world use water networks to distribute fresh water and remove sewage from private households and industrial facilities.
Their history reaches back into the dawn of civilization and they are generally understood as an essential constituent of our increasingly urbanized society.
Today, water networks are unanimously considered to be \emph{critical infrastructure}, because the lives and livelihood of urban populations directly depend on their functioning.
Due to the very high cost of directly accessing the infrastructure for upgrades and repair, pipes have been aging under the ground since they were first put into place decades or even centuries ago.
Meanwhile, the pressures excerted on the systems from extreme weather events such as prolonged dry periods and flash floods present an increasing threat to their seamless operation~\cite{fortier2015}.
For example, storm drains in the streets can clog, causing the streets to be flooded as seen during the heavy rain events in Berlin in 2017 and Washington DC in 2019. %
Furthermore, pipes can crack and the water leak out into the ground, which can cause significant harm to the surrounding infrastructures as well as economic damage due to the loss of fresh water.
Finally, in the case of a storm event, the load on the network can exceed its capacity and spill untreated into nearby waterways.
Such problems could likely be alleviated through large scale real-time measurements of key observables such as pressures, flow rates, temperatures~\cite{tsakalides2018}.
This information can help to determine the areas of greatest concern, schedule predictive maintenance, and ultimately control the various pumps and reservoirs more optimally.
\glspl{wsn} as powered by \gls{iot} technology present a promising tool for this purpose, because they can be deployed in parts of the network that are inaccessible to humans, stay operational over long periods without external power supply and transmit their sensor readings wirelessly to the cloud for further analysis.
In this paper, we report on \gls{wgs}, a joint project involving academic, industrial and municipal partners, that aims to develop an integrated solution based on power- and cost-efficient sensor devices as well as a scalable and fault tolerant data analytics platform.
The main contributions of our work can be summarized as follows:
\begin{enumerate}
    \item We discuss how the integration of the architecture components is subject to a tradeoff between protocol compliance and scalability.
    \item We propose an efficient scheme for enriching the stream of sensor measurments with values from a changing set of attributes and evaluate its throughput and latency for different cluster sizes and number of attributes. 
    \item We report a series of key insights from working on the algorithms addressing three of the many use cases in the WaterGridSense4.0 project.
    \item We provide all code required for deploying our platform in kubernetes with configurable cluster size, so our solution can be rolled out for arbitrarily large water networks around the world.
\end{enumerate}

The rest of this paper is structured as follows.
In the following section, we review previous efforts to leverage \gls{iot} technology for smart water grid monitoring.
Section~\ref{sec:arch}, describes our system architecture. %
The subsequent Section~\ref{sec:performance} reports on measurements with respect to several non-functional requirements.
Finally, we discuss uses cases and relevant algorithms in Section~\ref{sec:analytics} and conclude our work in Section~\ref{sec:conclusion}.

\section{Related Work}\label{sec:related}

With the recent proliferation of small, inexpensive wireless sensor devices with long battery life, a wide range of research projects have been initiated on the potential of~\gls{iot} technologies for various tasks in urban environments~\cite{rashid2016}.
Many focus specifically on monitoring and control of water grids~\cite{dong2015}, a direction in the literature often referred to as \emph{smart water grid}.
We take a look at the state of the art in this field throughout the remainder of this chapter.
It should be noted that we only consider recent approaches based on \glspl{wsn} technology to keep the discussion within the scope of our own work.\\

Among the seminal works on this subject is the \emph{PipeNet} system by Stoianov et al.~\cite{stoianov2007}.
They discuss an end-to-end architecture for \gls{wsn}-based pipe monitoring and leakage detection based on the Intel Mote platform.
The end device with multiple sensors is installed in a manhole and transmits the readings to the nearest gateway via Bluetooth which in turn relays them to an analytics backend via GPRS.
Despite their results being very relevant for our work, their research is a bit dated and uses a completely different technology stack from what is available today.
Similar approaches that rely on stationary sensors installed inside or around pipes include \emph{WaterWiSe@SG}~\cite{whittle2010}, \emph{MISE-PIPE}~\cite{sun2011}, and \emph{SWATS}~\cite{yoon2011}.
With the exception of MISE-PIPE, they all rely on some form of wireless transmission and focus on the development of the end device with the aim of detecting leaks and blockages.
Other publications focus more on the analysis of the data acquired from various meters installed across the district to detect anomalous behavior inside the local grid~\cite{perelman2012,loureiro2016}.

Another corpus of works considers mobile sensor nodes that move either actively or float passively through the pipes.
Among the earliest systems developed in this context is the modular \emph{MAKRO} robot for pipe inspection developed by Scholl et al.~\cite{scholl2000}.
It moves actively through the network and stores the recorded data locally to be retrieved via cable transfer after the robot has been extracted.
Similar principles are applied by the \emph{KANTARO} probe~\cite{nassiraei2007} that combines data from multiple sensors to obtain an even more detailed picture of the pipe condition.
Examples of passively floating probes include the~\emph{SewerSnort} system for gas monitoring~\cite{kim2009} and the damage-detection probe \emph{SPAMMS}~\cite{kim2010}.
All store their recorded observations locally (no wireless transmission) but differ in their approaches to in-pipe localization, a topic that we touch upon in Section~\ref{sec:leak} in our paper.
A more advanced scheme is presented in~\cite{lai2012}, where the authors propose \emph{TriopusNet}, a swarm-like \gls{wsn} consisting of multiple probes that autonomously (re-)position inside the pipe network.
They are mainly concerned with the problems of node placement and data routing.\\

The plethora of successful deployments of~\gls{wsn} for water network monitoring shows that the approach holds great potential for improved monitoring and control of water grids. %
However, they all use just a few nodes while a real deployment in a large city would need to integrate the data from thousands of nodes to provide a complete understanding of the network state.
Yet, real applications of~\gls{wsn} in critical infrastructures such as water grids are subject to stringent system requirements and introduce practical challenges such as data enrichment and protocol compliance.
Exactly these considerations are at the core of our research. %

\section{System Architecture}\label{sec:arch}

\begin{figure}
  \centering
    \includegraphics[width=0.98\columnwidth]{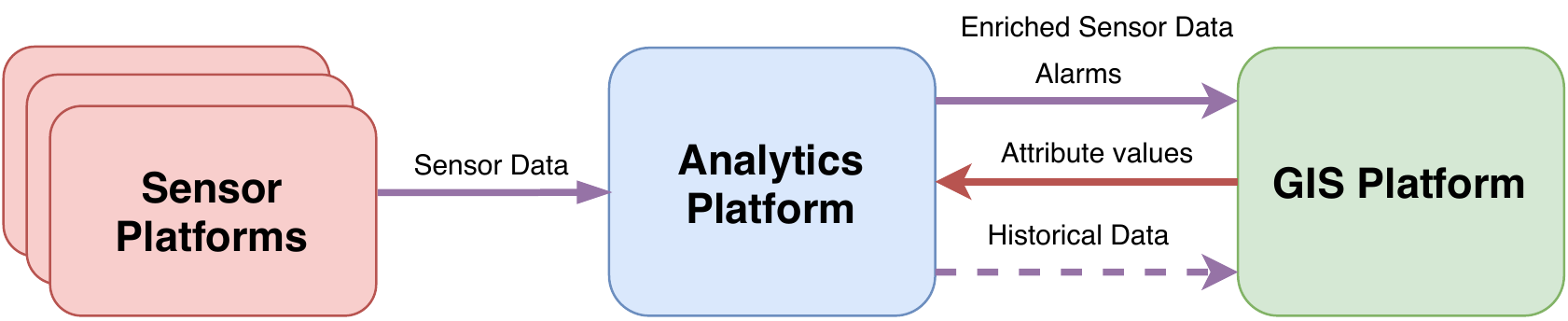}
    \caption{WaterGridSense4.0 system architecture. Solid lines represent data streams, dashed lines indicate batch data queries.}
    \label{fig:arch}
\end{figure}

The central problem addressed in the \gls{wgs} project is the continuous monitoring and analysis of large scale water infrastructures.
To that end, we develop a scalable and dependable cloud-based analytics platform that processes sensor measurements in real time.
The overall system architecture of \gls{wgs} is displayed in Figure~\ref{fig:arch} and is composed of three main parts:\\

The \emph{Sensor Platforms} control the sensor nodes that are placed in various locations across the water grid.
Their main tasks include power management, recording measurements, and sending them wirelessly to the Analytics Platform.
The sensor device developed within the \gls{wgs} project are modular, water-proof and can be deployed both as stationary sensors in key locations and as mobile devices floating through the pipes.

The \emph{Analytics Platform} is the central data processing component that is at the core of our research.
It receives a stream of measurements from the Sensor Platforms, processes the data and forwards the results to the GIS Plattform.

The \emph{GIS Platform} represents the point of interaction for the grid operators.
It receives the streams of processed sensor events and visualizes them on a map of the water network.
Additionally, the operators use the interface to send updates for the enrichment attributes to the Analytics Platform and query archived measurement series.\\

\subsection{The Analytics Platform}

The Analytics Platform is the central data processing component within the \gls{wgs} architecture.
It is composed of a set of interconnected services that are deployed using parametrized helm charts, so the system can be rolled out at arbitrary scale.
We made the code used for deployment of our plattform available online as open source\footnote{\url{https://github.com/dos-group/water-analytics-cluster}}.\\ %

\begin{figure}[t!]
  \centering
    \includegraphics[width=0.98\columnwidth]{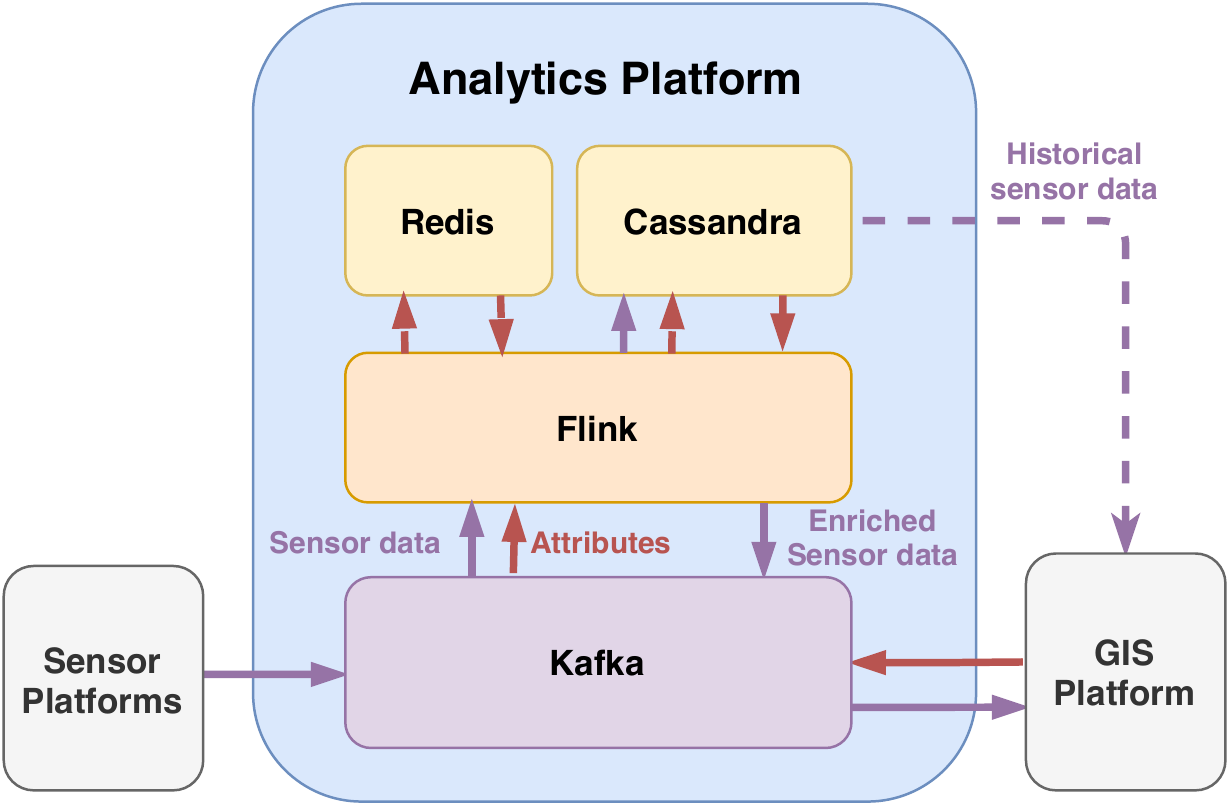}
    \caption{Architecture of the Analytics Platform.}
    \label{fig:analytics-arch}
\end{figure}

A complete schematic of the Analytics Platform is given in Figure~\ref{fig:analytics-arch} and can be summarized as follows:
The sensor measurements arrive at the platform through its data stream interface and are picked up by the distributed stream processing engine Apache Flink\footnote{\url{https://flink.apache.org/}}.
Flink then processes the data by subjecting it to the pipelines discussed in Sections~\ref{sec:enrichment} and~\ref{sec:analytics}.
The results of the analysis are again streamed to the GIS Platform for live visual display.
Additionally, all data is archived in the scalable distributed database Apache Cassandra\footnote{\url{https://cassandra.apache.org/}}.
The choice and configuration of the individual services is driven by two essential specifications: Requirements and protocols.\\

\paragraph{Requirements}
The requirements that our Analytics Platform must meet are defined by the water network operators which in turn must comply with municipal or federal regulations.
First and foremost, the system must be \emph{dependable}, i.e. allow continuous operations and produce reliable results even in the presence of partial system failures.
We address this issue by introducing service redundancy and using technologies that provide~\emph{exactly-once} semantics.
Through increasing redundancy, one can achieve almost arbitrary degrees of availability because it is increasingly unlikely that enough workers fail simultaneously to interrupt the service.
The second requirement concerns the \emph{responsiveness} of our system, i.e. its end-to-end latency: %
There should be a low delay between a measurement arriving at the data stream interface and the result of its analysis being sent to the GIS Plattform.
Obviously, this delay depends on the data processing jobs, as evaluated in Section~\ref{sec:performance}.
Furthermore, in order to apply the \gls{wgs} architecture to water networks of different sizes and with varying measurement frequencies, the Analytics Platform must provide \emph{scalability}.
To meet this requirement, the deployed services must support increasing the number of worker nodes and the algorithms used in the data processing pipelines must make use of the additional parallelism.
Finally, since we want our system to be freely available for municipalities around the world, the developed system must only consist of \gls{foss}.\\

\paragraph{Integration}
Unexpectedly, integration turns out to be a major challenge for the implementation of our Analytics Platform.
That is, the data stream interface connecting the three platforms must be at the same time fast, scalable, and compliant with the protocols used by the transmission infrastructure.
Within \gls{wgs}, LoRa has been determined as the most readily available wireless technology among the partnering cities.
As is typical in the \gls{iot} domain, the various LoRa implementations used by the infrastructure providers in the \gls{wgs} project use the \gls{mqtt} protocol for passing data upstream.
Among the most popular message broker implementations with \gls{mqtt} support is RabbitMQ\footnote{\url{https://www.rabbitmq.com/}}, which has been shown to achieve up to 40.000 packets per second throughput with a single node~\cite{dobbelaere2017}.
Its scalability is heavily limited by the fact that it does not allow the streams to be partitioned for parallel processing within Apache Flink.
On the other hand, Apache Kafka\footnote{\url{https://kafka.apache.org/}} is the fastest event streaming system available today with up to 420.000 packets per second~\cite{hesse2020} across partitioned streams but it does not natively support \gls{mqtt}.

Compromises in this tradeoff include using the \gls{mqtt} connector together with an external broker, relying on the \gls{mqtt} proxy plugin shipped with the enterprise version of the confluent platform, or, as of recently, using waterstream\footnote{\url{https://waterstream.io/}}, which merges the two technologies.
Unfortunately, the latter two options are not openly available and the first approach introduces a bottleneck, since the \gls{mqtt} broker has a significantly lower bound on its throughput, as stated above.
Within the project, we decided to use Kafka where supported by the local LoRa stack and additionally deploy RabbitMQ to provide MQTT compatibility. 
Throughout the remainder of this paper, we restrict our evaluation to Apache Kafka since it comes with scalable partitioning of the data streams.

\subsection{Data Enrichment}\label{sec:enrichment}

An important processing step that applies to all considered use cases is \emph{data enrichment}, i.e. the augmentation of the received sensor data with additional information.
For example, such information includes geolocations, device type information and measurement units which are not transmitted by the sensors in order to reduce energy consumption.
Attribute values are announced and updated via a separate Kafka data stream and stored within the Analytics Platform for use in the enrichment procedure.

Apache Flink supports stateful stream operations with RocksDB as a persistent key/value store, thus enabling fault tolerant~\emph{exactly-once semantics}.
Since values stored in Flink managed state can reside on disk instead of memory, we keep them in Java local memory but also store them in Redis\footnote{\url{https://redis.io/}} and Cassandra.
In case of a taskmanager unexpectedly dies, it will load the last checkpoint, replay the data stream and initialize its local state by obtaining the current value for each attribute from Redis.

\begin{figure}
  \centering
    \includegraphics[width=0.55\columnwidth]{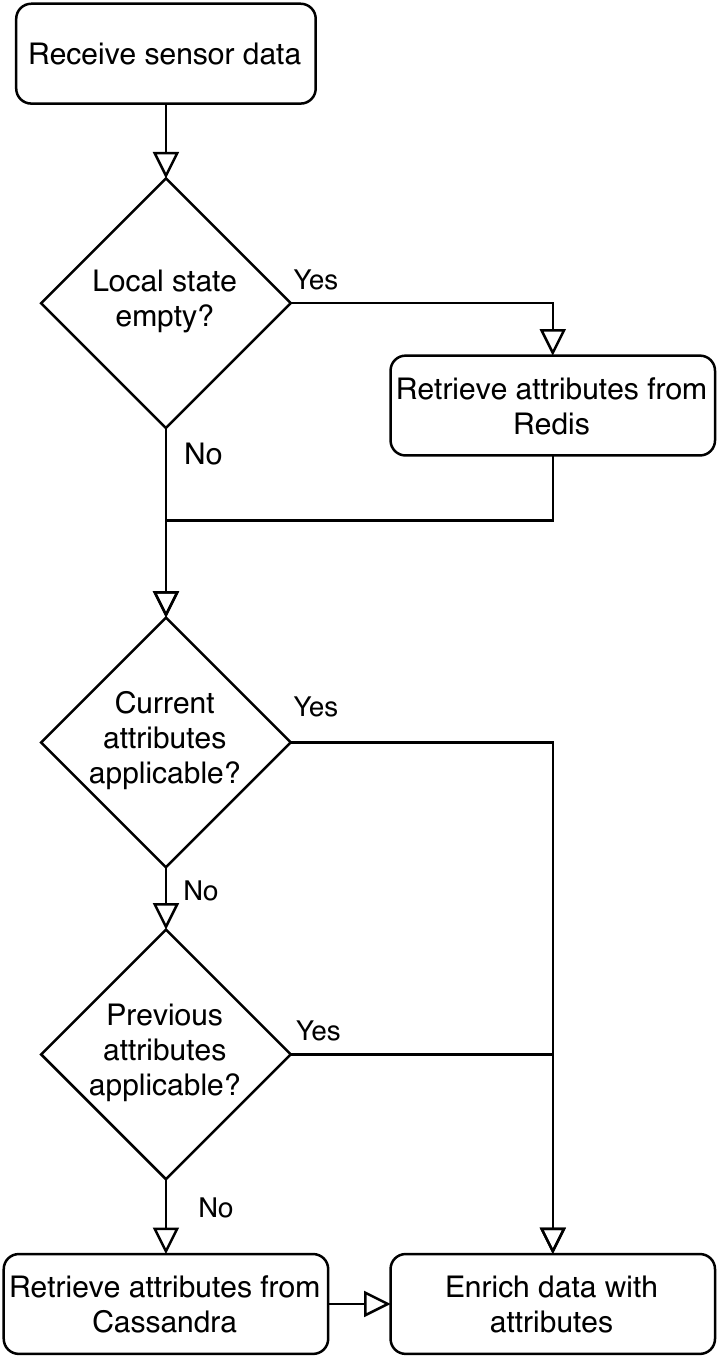}
    \caption{The complete data enrichment procedure.}
    \label{fig:enrichment}
\end{figure}

Due to the high volume of the stream and the lossy nature of LoRa networks, sensor data can arrive delayed and out-of-order.
Changes to the enrichment attributes on the other hand are comparatively rare and use reliable transmission which can cause the timestamps of the messages in the sensor data stream to lag behind those of the (already processed) attribute updates. %
In such cases, the enrichment procedure requires access to previously commissioned attributes to guarantee enrichment with information that was valid at the time when the measurement was taken.
At times of high load, this can happen for many measurements, which is why we also store the previous value for each attribute in local memory to avoid costly access to remote storage.
In case an out-of-order measurement arrives that is older than both the current and the previous value of a corresponding parameter, we retrieve the correct historical value from a distributed database.
The entire sequence of steps performed during data enrichment is presented in Figure~\ref{fig:enrichment}.
The complete source code is available in our git repository\footnote{\url{https://github.com/dos-group/water-analytics-enrichment}}.

\section{Performance Measurements}\label{sec:performance}

Our experimental setup consists of a 30-node Kubernetes cluster co-located with a HDFS cluster of the same size.
All nodes run on Intel Xeon E3-1230 V2 CPUs @ 3.30GHz with 16 GB RAM and 1 TB RAID0 HDD (linux sofware raid) and are connected via gigabit ethernet.
Each Flink job cluster consists of a single master in high-availability mode and three different cluster sizes taken from $\{4, 8, 12\}$.
All Flink workers are created with one task slot and 2 GB of memory.
A total of 3 runs of 30 minutes each were conducted for each job configuration. %

\begin{figure}[htp]
  \centering
    \includegraphics[width=0.8\columnwidth]{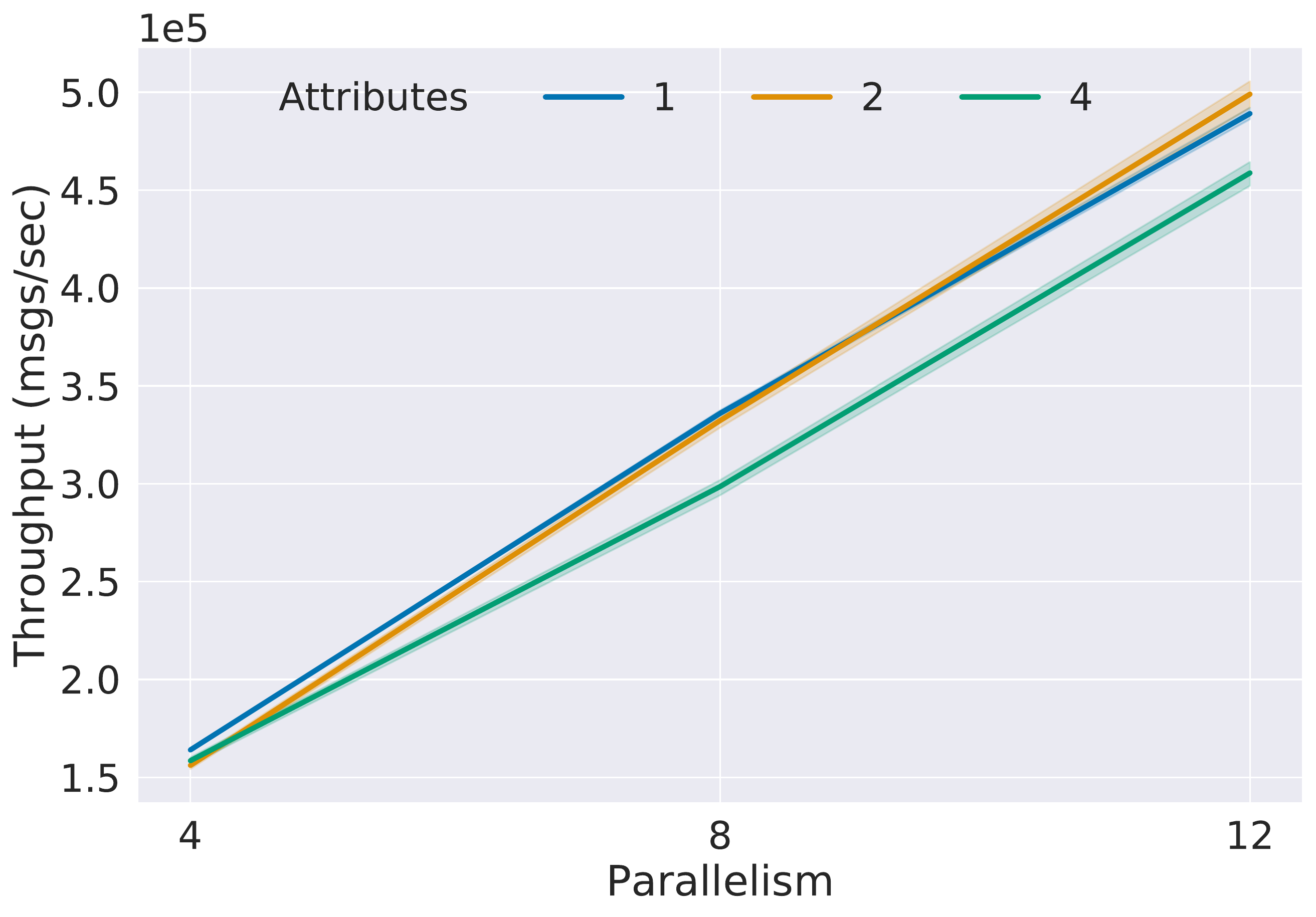}
    \caption{System throughput in mean number of messages per second for different cluster sizes and number of attributes.}
    \label{fig:throughput}
\end{figure}

The number of messages processed per second with our pipeline and system configurations are presented in Figure~\ref{fig:throughput}.
The results show that the data enrichment job achieves stable maximum throughput rates of roughtly 150.000 measurements per 4 taskmanagers and scales linearly with the size of the cluster.
The influence that the number of enrichment attributes has on performance seems to become larger with increasing cluster size, which we explain by the non-deterministic partitioning between flink taskmanagers and the redis cluster.
Furthermore, the performance difference between enrichment with 1 and 2 attributes is smaller than the residual variance among different measurement runs.
Across all configurations, the time it took the operator to ``warm up'', i.e. fill its local memory with attribute values, was around 45 seconds.

Concerning the requirement of responsiveness, mean operator latencies slightly increase with cluster size, from a median of 151ms (cluster size 4) to 159ms (cluster size 8) and finally 168ms (cluster size 12).
While this should be further investigated when considering even larger clusters, we consider these results to be in the range of acceptable values.
Overall, our results give conclusive evidence that our implementation is scalable with respect to the rate of incoming sensor data as well as the number of enrichment attributes.
In practice, one should configure the system to be at around 50\% load during normal operations to use resource efficiently but also give some room for load peaks.
For example, a system with cluster size 4 is appropriate for a network with 75.000 sensors that each send one measurement per second.

\section{Data Analytics}\label{sec:analytics}

The flexible design of the sensor nodes developed within \gls{wgs} allows deployments across a wide range of locations inside the water infrastructure:
It can be attached inside street inlets and manholes but also used as a floating probe for in-pipe inspection tasks.
From the many use cases thus enabled, we use this section to present three particularly important ones, report on our investigation into suitable algorithms and discuss some open issues to be addressed during the remainder of the project.\\

\subsection{Detecting Clogged Storm Drains}\label{sec:trumme}

Storm drains periodically clog due to leaves and litter that are washed in with the rain, which necessitates periodic cleaning of the internal collection container.
Since some are more affected than others, the cost of this maintenance could be significantly reduced if one could remotely estimate the degree of clogging without sending a team for local inspection.
The \gls{wgs} sensor nodes are intended to be installed within the street inlets to measure the water and dirt levels inside the inlets sludge containment and to estimate the urgency of cleaning them.
The cause and impact of clogging on drainage performance has been explored thoroughly~\cite{gomez2013,rietveld2020} and spatiotemporal correlations among nearby street inlets have been shown to exist~\cite{pulido2019}.
Among the proposed solutions is a passive system for estimating water levels in runoffs using RFID tags~\cite{atojoko2013} and a zigbee-based \gls{wsn} using acoustic sensors~\cite{see2011}.

\subsection{Predicting Combined Sewer Overflows}\label{sec:cso}

In a combined sewer system, stormwater runoff runs through a single pipe together with wastewater from homes, businesses, and industry.
During periods with heavy rainfall, the amount of stormwater can become greater than the network capacity and cause a \gls{cso}. %
This problem could be alleviated by monitoringing the water network in order to predict imminent load spikes and initiate countermeasures, such as preemtive clearing of rainwater tanks and basins.
Previous research on \gls{cso} detection covers the use of neural networks and control limit theory to detect \gls{cso} events from flow measurement time series~\cite{sumer2007}.
In this work, the authors also propose a method to incorporate data from multiple geospatial locations to improve detection.
Sonnenberg et al.~\cite{sonnenberg2011} compare multiple approaches to \gls{cso} detection based on water level measurements, rainmeter data and even a simulation model of the water grid. %

\subsection{Detecting Leakages and Inflows}\label{sec:leak}

Presumably, the use case with the most existing works in the literature concerns the remote detection of leakages and inflow of rainwater or groundwater into wastewater pipes.
Many sensor-based solutions have been proposed for this problem, as discussed in Section~\ref{sec:related}.
To cover this use case within \gls{wgs}, sensor nodes are designed to be capable of floating through the pipes while continuously recording key observables, such as temperature and conductivity of the surrounding water.
Once they come within range of a gateway, they transmit their measurements to the Analytics Plattform for further inspection.
The main difficulty encountered in this use case comes from the fact that radionavigation systems such as GPS don't work deep in the ground and that therefore, the location of each measurement has to be estimated by some other means.
Recent solutions to this problem include acoustic localization~\cite{kurtz2006,kumar2017}, using gyroscope data~\cite{lai2012,zheng2018}, and RSSI-based localization with respect to anchor points, e.g. the above mentioned gateways~\cite{kim2009,gong2016}.

\subsection{Open Issues}

Thorough analysis of the previously proposed solutions for the three use cases reveals a set of required algorithmic features.
They include
\begin{itemize}
    \item the integration of weather data through a separate data stream,
    \item the support for geospatial queries, and finally
    \item a scalable solution for detecting anomalous values based on spatial correlations with neighboring sensors
\end{itemize}
The feasibility of including weather data depends on the products offered by the respective national weather service.
In the case of the German DWD, open weather data is made available in the form of daily reports and forecasts that have to be crawled and turned into a data stream to be used within our platform.
The second issue concerns the use of geospatial information, e.g. to process readings of physically co-located sensors together.
Redis supports geospatial queries by means of the \texttt{GEOHASH} and related primitives, but the frequent transmission of query results would likely degrade performance significantly as opposed to performing such computations locally. %
To fill this gap, the GeoBeam and GeoFlink frameworks have been introduced.
They extend the Flink API to enable geospatial queries~\cite{he2019,shaikh2020}.

Finally, a scalable time series anomaly detection method is needed for all considered use cases, which should provide a means to incorporate information about the geospatial placement of the sensors.
There are promising recent results describing a scalable \gls{arima} implementation in Flink for detecting anomalies in hydrologic time series~\cite{ye2020}.
Previously, the use of \gls{lisa} was proposed for anomaly detection in water networks~\cite{difallah2013}.
We are currently working on a pipeline that combines weather- and sensor data from across a certain area and applies spatio-temporal anomaly detection to identify use case-specific target events.

\section{Conclusion}\label{sec:conclusion}

Intelligent monitoring of large scale water infrastructures is an open problem addressed by the \gls{wgs} project.
As part of its system architecture, we introduce an openly available data analytics platform that runs in the cloud.
In this paper, we reported on key insights gained during the development process including component integration, platform design and data analysis.
Our platform is shown to meet important critical infrastructure requirements, specifically dependability, responsiveness and scalability.
We further described a complete data enrichment procedure optimized to fit the characteristics of the application domain.
In a series of experiments, we demonstrate scalability, quantify responsiveness and provide a point of reference for choosing the right cluster size in any given deployment scenario.
Throughout the remainder of the project, we plan to implement and evaluate the outlined analysis algorithms for which certain open issues need to be addressed.
These include the integration of weather data, the support for geospatial queries and the use of a scalable anomaly detection method in the stream processor.

\section*{Acknowledgments}

This work has been supported through a grant by the German Ministry for Education and Research (BMBF) as WaterGridSense 4.0. We would like to thank all partners involved in this project - Hamburg Wasser, Berliner Wasserbetriebe, Funke Group and ACO Severin Ahlmann GmbH \& Co. KG. %

\bibliographystyle{IEEEtran}
\bibliography{refs}

\end{document}